\documentclass[11pt]{article}
\usepackage[margin=1in]{geometry}
\usepackage{graphicx}
\usepackage{booktabs}
\usepackage{amsmath}
\usepackage[hidelinks]{hyperref}
\usepackage{xcolor}

\title{Selection, Not Salience:\\
The Shape and Limits of Personalization in Social Highlighting}
\author{
  Kazuki Nakayashiki \quad Keisuke Watanabe \\[3pt]
  Glasp Inc. \\
  \texttt{kazuki@glasp.co} \quad \texttt{kei@glasp.co} \\[3pt]
  {\small\itshape Co-first authors (equal contribution).}
}
\date{}

\begin{document}
\maketitle

\begin{abstract}
Does personalizing what a reader sees pay off, and where does it stop? Building on
a co-readership identity control --- the same document highlighted by many users,
which holds the document and topic fixed and asks whether a person's own history
predicts their marks better than another reader's does --- we map the shape and
limits of individual personalization in highlighting across reading altitudes. (1)
At the \emph{document} altitude we give the clean, leakage-free, identity-controlled
measurement that prior next-document evaluations could only upper-bound: a person's
history identifies which documents in a co-reading neighborhood are theirs, with an
own-versus-other gap of $+0.169$ against community negatives and $+0.119$ against
topic-matched hard negatives (both highly significant); a smaller content-based arm
suggests the signal is not purely title-driven but is largely thematic.
This is comparable to the span-level selection signal ($+0.14$) reported in our prior
work~\cite{nakayashiki2026salience}: the selection signal is of \emph{comparable
magnitude across altitudes} ($+0.12$ to $+0.17$, span and document alike), most of it
stable topic preference. (2) At the \emph{sentence} altitude, a two-stage personalized
auto-highlight --- an impersonal salience model proposes candidates, a personal model
re-ranks them --- does not improve on its impersonal baseline: two off-the-shelf
zero-shot LLMs (including a frontier model) predict highlight locations \emph{worse}
than a trivial lead baseline, and personal re-ranking is \emph{beaten} by the salience
order \emph{even on the highest-recall candidate pool}, indicating the null is not
merely a Stage-1 ceiling artifact. In our setting, measurable personalization appears primarily at the
\emph{selection} layer: modest ($\sim$$+0.13$), topic-dominated, with no reliable gain
from pushing it into the sentence-level salience layer. Along the way we surface a
methodological hazard --- a control-in-negatives bias that inflated our own document gap
to a spurious $+0.227$ until audited --- and connect our findings to the
recommender-systems result that popularity is hard to beat and the
summarization-evaluation result that non-personalized models often beat personalized
ones. The results suggest that going beyond the shared salience layer is better
approached by \emph{aggregating} individuals than by personalizing them harder.
\end{abstract}

\section{Introduction}
A recurring premise behind digital twins and generative agents is that a person's
behavioral traces reveal their preferences and predict their choices~\cite{park2024,
salemi2023}. Highlighting is an unusually rich such trace. In prior work we asked
how much of an individual is recoverable from highlighting and found a sharp
\emph{asymmetry}: which sentences a person marks within a document is mostly
\emph{shared salience} (the crowd dominates; the personal residual is a whisper,
own-versus-other gap $+0.017$ average precision), whereas which of the
already-salient passages are \emph{theirs} carries a real individual signal
($+0.14$), most of it stable topic preference~\cite{nakayashiki2026salience}.

That paper located \emph{where} individuality lives. This paper asks two questions
about its \emph{shape and limits}, both of which matter for whether personalizing a
reader's experience is worth doing. First, does the selection signal hold at a
\emph{coarser} altitude --- not which span within a document, but which
\emph{document} a person chooses --- and how large is it once the confounds that
plague next-document evaluations are removed? Second, can personalization be pushed
\emph{into} the salience layer: if an impersonal model hands a reader a small set of
generically highlight-worthy candidates, does a personal model usefully re-rank them?

We answer both with the same co-readership identity control, lifted to each altitude.
Our findings are: (i) the document-selection signal is real, leakage-free, and
identity-controlled at $+0.12$ to $+0.17$ --- the same ballpark as span-level
selection, so the selection signal is of comparable magnitude across altitudes; and (ii)
sentence-level personalization does \emph{not} improve on the impersonal salience order
--- a result robust to the Stage-1 recall ceiling --- because that order is built on a
shared salience layer that an off-the-shelf LLM models worse than position and that
personal re-ranking does not beat. Together these map a boundary: on this
social-highlighting data, measurable personalization is a selection-level,
topic-dominated, modest effect for which we find no benefit at the salience layer.

\paragraph{Contributions.}
(1) A clean, leakage-free, \emph{identity-controlled} measurement of document-level
selection ($+0.169$ community / $+0.119$ topic-matched), correcting the upper-bound
status of prior next-document results, and the observation that the selection signal
is of comparable magnitude across altitudes ($+0.12$--$0.17$, spans and documents
alike). (2) A negative
result: a two-stage personalized auto-highlight does not beat its impersonal
baseline, and two off-the-shelf zero-shot LLMs, including a frontier model, underperform
a lead heuristic as salience models. (3) A methodological caution we found by auditing our own result:
placing identity-control users' documents among the candidate negatives inflates the
gap (here from a clean $+0.169$ to a spurious $+0.227$); we report the symptom and the
fix. (4) A synthesis that connects reading personalization to the
recommender-systems ``popularity is hard to beat''~\cite{krichene2020} and the
summarization ``non-personalized beats personalized''~\cite{perseval2024} literatures,
and motivates aggregation over personalization.

All estimates are leakage-free, use a method-matched own-versus-other identity control,
and carry 95\% cluster-bootstrap confidence intervals by document and by
user.\footnote{The paper and figures are available at
\url{https://github.com/glasp-co/personal-selection}. The scoring pipeline and per-pair
results run against private user highlighting behavior and are not released; the
cluster-bootstrap estimator and aggregate statistics are available to researchers on
reasonable request.}

\section{Related work}
\paragraph{Personalized highlighting and summarization.}
Personalized highlight detection has been studied mostly in video, conditioning a
ranker on a user's prior selections~\cite{gygli2018}; the personalized-summarization
line conditions extractive or abstractive models on user
profiles~\cite{salemi2023,ao2021}. Our two-stage architecture (an extractive
candidate pool re-ranked by a personal model) is not new; our contribution is the
\emph{measurement} --- with a real-behavior identity control --- of whether it pays
off, rather than a new model.

\paragraph{Is personalization overclaimed?}
Two literatures independently caution that personal models are weaker than they look.
In recommendation, an unpersonalized \emph{popularity} baseline is empirically hard to
beat and sampled-negative evaluations can distort comparisons~\cite{krichene2020,
revisitpop2020}. In summarization evaluation, the EGISES / PerSEval line finds that
non-personalized models often score above personalized ones~\cite{vansh2023,
perseval2024}. Our results are the highlighting instance of the same phenomenon, made
precise with a co-readership control.

\paragraph{LLMs and salience.}
LLMs summarize well, yet their internal notion of salience correlates only weakly with
human salience~\cite{trienes2025}; our two-stage result is the behavioral counterpart
(a single LLM predicts human highlight locations below a positional baseline).
Highlighting has long been understood as partly social and partly idiosyncratic in the
reading and education literatures~\cite{winchell2020}; we quantify both sides.

\paragraph{Aggregating individual models.}
Ensembles of \emph{diverse} models can rival human crowds at
forecasting~\cite{schoenegger2024}. This motivates, but does not answer, the question
we leave to future work: whether many behaviorally-grounded reader models, aggregated,
can reconstruct collective salience.

\section{Data and method}
Glasp is a social web highlighter; for documents marked by multiple users we know
exactly who highlighted what, materializing a \emph{co-readership} structure rarely
available in personalization research. This lets us build a method-matched
\emph{identity control}: for a target user $A$, we compare a scorer driven by $A$'s
own history (\textsc{own}) against the identical scorer driven by a comparable other
reader $B$'s history (\textsc{other}), on the \emph{same} candidate set; the paired
$\textsc{own}-\textsc{other}$ gap is robust to whatever makes the candidate set easy or
hard, because the conditioning cancels. We report this gap, plus contrasts against a
topically-nearest peer (\textsc{near}), the crowd, and a random floor.

Profiles are \emph{leakage-free}: a user's profile excludes every candidate being
scored (and near-duplicate documents), and we additionally compute a \emph{leaked}
profile (candidates included) to quantify inflation. Average precision (AP) is the
per-instance metric; all intervals are 95\% cluster bootstraps resampling whole
documents (primary) and whole users (robustness). We study two altitudes beyond the
within-document span level of~\cite{nakayashiki2026salience}: \textbf{document
selection} (\S\ref{sec:docsel}) and \textbf{sentence-level two-stage auto-highlight}
(\S\ref{sec:twostage}). Table~\ref{tab:data} summarizes the two datasets; further setup
details are in Appendix~\ref{app:repro}.

\begin{table}[t]
\centering
\small
\begin{tabular}{lcc}
\toprule
 & Exp.~A (documents) & Exp.~B (sentences) \\
\midrule
(user, item) pairs            & 519 & 1{,}011 \\
distinct users                & 489 & 867 \\
documents / communities       & 94 communities & 60 documents \\
candidates per pool (median)  & 30 & $K$ (5--20) \\
positives per pair (median)   & 7 & --- \\
negatives per pair (median)   & 25 & --- \\
mean base rate                & 0.23 & --- \\
\bottomrule
\end{tabular}
\caption{Dataset and pool statistics. Embedding model
\texttt{text-embedding-3-small} (512-d); inference by $4{,}000$-iteration cluster
bootstrap (by document and by user).}
\label{tab:data}
\end{table}

\section{Experiment A: document selection}
\label{sec:docsel}
\paragraph{Setup.}
We sample seed documents with several co-readers; the co-readers form a community. For
a target member $A$ we time-split $A$'s documents (older $=$ profile, newest $30\%$ $=$
held-out positives), so the profile is disjoint from the candidate pool by
construction. \emph{Negatives} are documents the \emph{other} community members
highlighted but $A$ did not --- the same interest ecosystem, so the test is not the
trivial one of distinguishing $A$'s documents from a random global pool. The candidate
pool is $A$'s held-out positives plus these co-reader negatives; each document is
represented by its title embedding, each profile by the mean embedding of its
highlighted spans. We score the pool by \textsc{own} (cosine to $A$'s profile),
\textsc{other} ($B$'s profile, $B$ the most-prolific other member), \textsc{near}
(the topically-closest member), the within-pool crowd (owner count), and a random
floor, under two negative regimes: \emph{community} (a broad sample) and \emph{hard}
(the co-reader documents most similar to $A$'s profile --- the strict within-topic
test). Final sample: $519$ (user, community) pairs over $489$ users and $94$
communities.

\paragraph{A bias we had to fix.}
Our first run reported an own-versus-other gap of $+0.227$. Auditing it, we found that
the identity-control members' \emph{own} documents were sitting in $A$'s negatives
(they are ``other members' documents $A$ did not highlight''); scoring the pool by a
control's profile then ranks that control's own documents --- labelled negative --- to
the top, dragging $\textsc{other}$ and $\textsc{near}$ \emph{below} the random floor
(Table~\ref{tab:docsel}) and inflating the gap. Excluding the control members'
documents from the negatives (and adding a near-duplicate guard on the profile)
removes the artifact and yields the numbers we report. We flag this because it is an
easy and silent mistake in any co-readership identity control: the diagnostic is a
control scoring below random.

\paragraph{Results.}
Table~\ref{tab:docsel} gives the corrected estimates. The clean identity gap is
$+0.169$ (community, 95\% CI $[0.141, 0.199]$) and $+0.119$ (hard, $[0.094, 0.146]$),
with every document- and user-bootstrap positive. Against the conservative
topically-nearest peer it is $+0.141$ (community) / $+0.112$ (hard); against chance,
$+0.220$ / $+0.092$. About a third of the community gap is coarse topic (community
minus hard $\approx +0.05$); the rest survives topic-matched negatives. The within-pool
crowd is near or below chance (below the random floor), but here this is largely
\emph{mechanical} --- a user's newest documents have accrued fewer co-readers --- so,
unlike the clean span-level result, we do not read it as evidence that popularity is
anti-diagnostic.

\begin{table}[t]
\centering
\small
\begin{tabular}{lcc}
\toprule
Scorer (mean AP) & community neg. & hard neg. \\
\midrule
\textsc{own} (A's history)      & \textbf{0.529} & \textbf{0.411} \\
\textsc{near} (closest peer)    & 0.388 & 0.299 \\
\textsc{other} (control $B$)    & 0.360 & 0.292 \\
random floor                    & 0.310 & 0.320 \\
crowd (in-pool popularity)      & 0.228 & 0.224 \\
\emph{leaked} profile           & 0.565 & 0.448 \\
base rate                       & 0.229 & --- \\
\midrule
$\textsc{own}-\textsc{other}$ (identity gap) & $+0.169$ & $+0.119$ \\
\quad 95\% CI (by document)     & $[0.141,0.199]$ & $[0.094,0.146]$ \\
$\textsc{own}-\textsc{near}$    & $+0.141$ & $+0.112$ \\
$\textsc{own}-$ random          & $+0.220$ & $+0.092$ \\
\bottomrule
\end{tabular}
\caption{Document selection ($519$ pairs, $489$ users, $94$ communities). After
removing the control-in-negatives bias, the controls (\textsc{other}, \textsc{near})
sit \emph{above} the random floor in the community regime, as a fair control must; the
identity gap is $+0.169$ / $+0.119$, every bootstrap positive.}
\label{tab:docsel}
\end{table}

\paragraph{Comparable across altitudes.}
The clean document gap ($+0.12$ to $+0.17$) is in the same range as the span-level
selection gap ($+0.14$) of~\cite{nakayashiki2026salience}, and the bootstrap intervals
broadly overlap. The selection signal is of \emph{comparable magnitude} across altitudes,
not larger at the document level --- contrary to our own first, biased estimate of
$+0.227$ --- and it remains topic-dominated (Figure~\ref{fig:altitude}).

\paragraph{Content representation (robustness).}
Because a title could encode topic without content, we re-ran a subset with each
candidate represented by its \emph{article-content} centroid (the mean embedding of its
first sentences) rather than its title; article fetching limits this arm to $N=42$ pairs
over $16$ communities. A reader's content profile predicts their held-out documents well
above chance --- $\textsc{own}-$ random $=+0.154$ against topic-matched \emph{hard}
negatives (CI $[0.043, 0.283]$, every bootstrap positive) --- so the document signal is
not \emph{purely} a title artifact. The own-versus-other identity gap stays positive
($+0.05$ community, $+0.03$ hard) but is not significant at this small $N$, and it
compresses because content embeddings make \emph{any} reader's profile a decent
predictor in a topically homophilous co-reading neighborhood (the control
$\textsc{other}$ rises from $0.36$ in the title arm to $0.50$ here). We therefore keep the well-powered title-based gap as the headline and read the
document-altitude signal as content-grounded but, like span-level selection, largely
thematic.

\section{Experiment B: sentence-level two-stage auto-highlight}
\label{sec:twostage}
\paragraph{Setup.}
The product motivation is a personalized auto-highlight that works on a \emph{cold}
document: an impersonal Stage~1 proposes $K$ candidate sentences, a personal Stage~2
re-ranks them to surface the few a given user would mark. We validate this offline on
co-read documents (about $1{,}000$ (user, document) pairs), with a strict
own-versus-other control. Stage~1 generators are an
\textbf{LLM} (``select the sentences a reader would highlight,'' no user information),
\textbf{generic} centrality, \textbf{lead} (document order), the \textbf{crowd} union,
and \textbf{random}; Stage~2 re-ranks by cosine between each candidate sentence
embedding and the user's leakage-free highlight-profile centroid (own), versus another
reader's centroid (other). We measure Stage-1 \textit{recall@K} (does the pool even
contain the user's marks?) and the within-pool personal re-rank.

\paragraph{Results.}
Table~\ref{tab:twostage} reports recall@$10$. An off-the-shelf zero-shot LLM Stage~1
recovers the user's highlights \emph{worse} than the trivial lead baseline and far
below the crowd --- and a frontier model (\texttt{gpt-5.5}) barely improves on a small
one (\texttt{gpt-4o-mini}), consistent with LLM salience correlating only weakly with
human salience~\cite{trienes2025}. Within the LLM pool the own-versus-other gap is a
small $+0.015$ (95\% CI $[-0.007, 0.038]$); the personal re-rank \emph{loses} to
Stage-1's own salience order ($\textsc{own}-\textsc{stage1} = -0.05$, CI
$[-0.092, -0.008]$).

\paragraph{The null is not a Stage-1 ceiling artifact.}
One might worry that the LLM pool's modest recall ($0.25$) leaves Stage~2 no room: a
re-ranker cannot recover marks the pool never contained. We therefore repeat the
within-pool re-rank on the \emph{highest}-recall pool, the crowd union (recall@10
$=0.42$, recall@20 $=0.58$). Personal re-ranking still \emph{loses} decisively to the
salience order there --- here the crowd's popularity order --- by $-0.118$ at $K{=}10$
and $-0.135$ at $K{=}20$ (\emph{every} bootstrap negative). If anything, raising recall
\emph{enlarges} the loss ($-0.05$ on the lower-recall LLM pool versus $-0.118$ and
$-0.135$ here) --- the opposite of what a Stage-1 ceiling would produce, where more
recall should give Stage~2 more room. A small own-versus-other signal does exist on that
pool (identity gap
$+0.022$, CI $[0.002, 0.042]$, at $K{=}20$) --- personalization is not \emph{nothing} ---
but it is smaller than the advantage of simply trusting the impersonal salience order.
Across pools and models we find no setting in which personal re-ranking improves on
that order at the sentence level.

\begin{table}[t]
\centering
\small
\begin{tabular}{lcc}
\toprule
Stage-1 generator & recall@10 (\texttt{4o-mini} run) & recall@10 (\texttt{gpt-5.5} run) \\
\midrule
crowd union              & 0.42 & 0.41 \\
lead (document order)    & 0.29 & 0.30 \\
generic centrality       & 0.22 & 0.26 \\
\textbf{LLM}             & \textbf{0.22} & \textbf{0.25} \\
random                   & 0.19 & 0.20 \\
\bottomrule
\end{tabular}
\caption{Two-stage Stage-1 quality. The LLM ($\approx 0.22$--$0.25$) recovers the
user's highlights below the lead heuristic ($0.29$--$0.30$) and far below the crowd
($0.41$--$0.42$); a frontier model does not change this. Within-pool personal re-rank
identity gap: $+0.015$ (n.s.); personal $-$ salience-order: $-0.05$.}
\label{tab:twostage}
\end{table}

\section{Synthesis: the shape and limits of personalization}
\begin{figure}[t]
\centering
\includegraphics[width=0.82\linewidth]{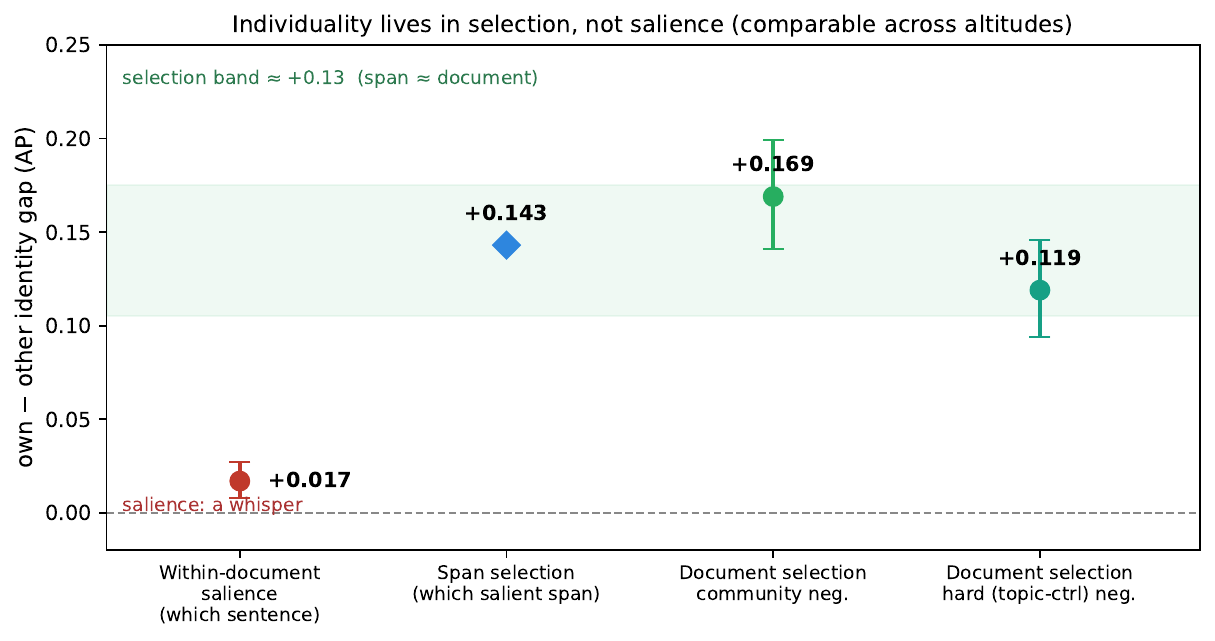}
\caption{The own-versus-other identity gap by reading altitude (Paper~I values for
salience and span selection; this paper for document selection). Individuality is a
whisper in salience ($+0.017$) and a consistent, modest signal in selection
($\sim$$+0.13$) whether the unit is a span or a document. The three altitudes use
different candidate constructions and base rates, so this is a qualitative comparison
(same ballpark), not a single like-for-like metric (\S\ref{sec:limits}).}
\label{fig:altitude}
\end{figure}

Figure~\ref{fig:altitude} draws the picture. In our social-highlighting data, the
measurable personal signal is a \emph{selection} phenomenon: it is near-zero in salience
(which sentence is important)
and a consistent $\sim$$+0.13$ in selection (which of the salient things are yours),
whether the things are spans or documents. It is modest and topic-dominated, and ---
Experiment~B --- we find no benefit from pushing it into the salience layer, because
that layer is shared, is modeled worse than position by an off-the-shelf LLM, and is not
improved by personal re-ranking (even on the highest-recall pool). The crowd's role is
altitude-dependent in our own data, which is itself informative: among already-salient
\emph{spans} (Paper~I) and among a user's newest \emph{documents} (Experiment~A)
popularity is near or below chance, yet it is the \emph{strongest} sentence-level
Stage-1 generator (Table~\ref{tab:twostage}). Popularity is diagnostic of what is
salient but anti-diagnostic of which salient thing is an individual's. This is the
highlighting instance of two findings reached independently elsewhere: an unpersonalized
popularity baseline is hard to beat in recommendation~\cite{krichene2020,
revisitpop2020}, and non-personalized models often beat personalized ones in
summarization~\cite{vansh2023, perseval2024}.

\section{Discussion}
For products, the boundary is actionable. A generic auto-highlight (positional or
crowd-transfer salience) is the right tool at the sentence level; personalization buys
little there and can hurt. Where personalization pays is \emph{selection} --- which
documents, sections, or topics to surface to a given reader --- and even there the
signal is modest and largely thematic, so it is best understood as interest routing,
not a deep behavioral fingerprint.

The more interesting implication is for \emph{collective} salience. Because individual
reading is mostly shared salience plus a thin idiosyncratic layer, the way to exceed a
single reader is not to personalize harder but to \emph{aggregate} many readers. In this
zero-shot extractive setup, off-the-shelf LLMs are weak models of human salience;
whether many \emph{behaviorally-grounded}
reader models, aggregated, can reconstruct --- or decompose into sub-populations --- the
collective Popular Highlights of a document is an open question that our identity
control and data are well suited to answer. We leave it to future work.

\section{Limitations}
\label{sec:limits}
\textbf{Scope.} Our population is doubly self-selected --- people who install a social
highlighter, and documents popular enough to be co-read --- so we describe
personalization in \emph{social highlighting on a co-readership platform}, not reading
in general. \textbf{Altitude comparison.} The altitudes differ in candidate construction
and base rate, and an AP gap is not scale-free, so ``comparable across altitudes'' is a
qualitative statement; a base-rate-normalized re-analysis is left to future work.
\textbf{Hard negatives.} In Experiment~A, \textsc{hard} negatives are chosen by
similarity to $A$'s \emph{own} centroid, which is adversarial to \textsc{own} and
neutral to \textsc{other}; this makes the hard-negative gap ($+0.119$) a
\emph{conservative} estimate. \textbf{Personal re-ranker.} The Stage-2 personalizer is a
profile-centroid cosine scorer; a learned personal re-ranker is left to future work,
though the own-versus-other control bounds the personal information any scorer of this
family could exploit. \textbf{Other.} Negatives are implicit (``$A$ did not
highlight $X$'' $\neq$ ``$A$ would not''; the noise is symmetric across
\textsc{own}/\textsc{other}, so it does not bias the gap but caps absolute AP); the main
document representation is title-based, which we probe with a content-based robustness
arm (\S\ref{sec:docsel}); the within-pool crowd in Experiment~A is mechanically weak;
the $94$ communities are a modest number of high-level clusters; and we cannot cleanly
separate a non-thematic ``style'' residual from finer-grained topic, as in Paper~I.

\section{Conclusion}
On a social-highlighting platform, personalization has a clear shape and a clear limit.
Individuality surfaces in \emph{selection} --- which things a reader picks --- where it
is a consistent but modest, topic-dominated signal of about $+0.13$, of comparable
magnitude whether the unit is a span or a whole document. We find no corresponding
signal in \emph{salience}: a two-stage personalized auto-highlight does not beat its
impersonal baseline, and in our zero-shot extractive setup off-the-shelf LLMs, including
a frontier model, underperform a lead heuristic as salience models. These results suggest that going beyond the shared
salience layer is better approached by aggregating individuals than by personalizing
them harder.

\section*{Ethics and data}
The experiments use aggregate behavioral traces from a social-highlighting platform. We
report only aggregate statistics; per-user and per-pair data are not released because
they can reveal private reading behavior. Profiles are anonymized for analysis; the
scoring pipeline runs against private user data and is likewise not released, though the
cluster-bootstrap estimator and aggregate statistics are available to researchers on
reasonable request.

\appendix
\section{Reproducibility details}
\label{app:repro}
\textbf{Embeddings.} \texttt{text-embedding-3-small} (512-d) for all titles, profile
spans, and (the Experiment~A content arm) article-content sentences.
\textbf{LLMs (Experiment~B Stage~1).} \texttt{gpt-4o-mini} and \texttt{gpt-5.5} as
served by the OpenAI API in 2026-06 (no dated snapshot was pinned), zero-shot,
temperature~$0$. The prompt presents the document as numbered
sentences with \emph{no} user information and asks the model to return ``the sentence
numbers a typical reader would most likely highlight, most-likely first, at most $K$.''
We did not tune the prompt or use few-shot / chain-of-thought; the negative result is
for this off-the-shelf zero-shot setting and may not hold for heavily engineered
prompts. $K \in \{5,8,10,15,20\}$; the headline pool is $K=10$.
\textbf{Document selection (Experiment~A).} Seed documents have $\geq 8$ co-readers; per
target $A$, the newest $30\%$ of $A$'s documents are held-out positives and the older
documents the profile; up to $\sim$$5$ negatives per positive; \textsc{hard} negatives
are the top co-reader documents by cosine to $A$'s profile centroid; near-duplicate
profile documents (title cosine $\geq 0.92$ to a positive) are dropped; the
identity-control $B$ is the most-prolific other member, and $B$'s and the topic-near
member's own documents are excluded from $A$'s negatives (the control-in-negatives fix).
\textbf{Inference.} Average precision per pool; $4{,}000$-iteration cluster bootstrap
resampling whole documents (primary) and whole users (robustness). The scoring pipeline
runs against private user data and is not released; the cluster-bootstrap estimator and
aggregate statistics are available to researchers on reasonable request.

\end{document}